# Scientometrics and Information Retrieval - weak-links revitalized

**Philipp Mayr, Andrea Scharnhorst**

This special issue brings together papers from experts of communities which often have been perceived as different once: bibliometrics/scientometrics/informetrics on the one side and information retrieval on the other. Our motivation as guest editors started from the observation that main discourses in both fields are different, that communities are only partly overlapping[1] and from the belief that a knowledge transfer would be profitable for both sides. Hereby, we (see also Karlsson et al. in this issue) were inspired by the bibliometric analysis of the broader field of Library and Information Science done by White and McCain. (White and McCain 1998; Zhao and Strotmann 2014)

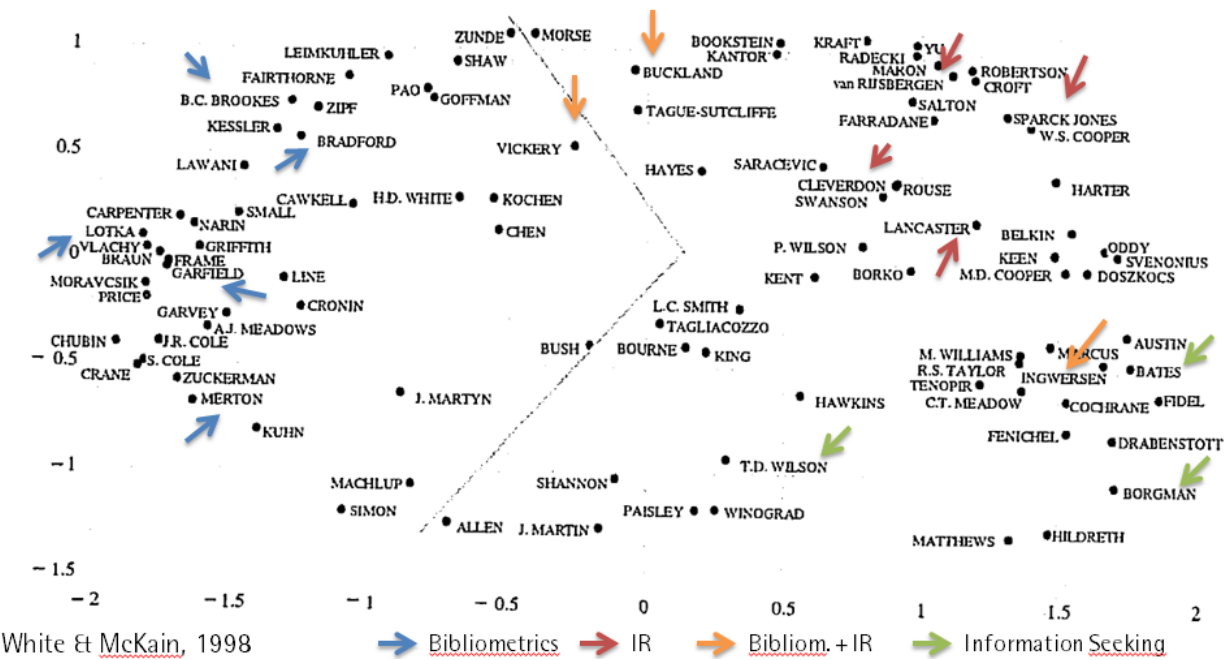

**Fig 1.** Bibliometric analysis of the broader field of Library and Information Science. Original from White and McCain (1998). Used at the introduction of the ISSI 2013 and ECIR 2014 workshop. See highlighted authors on the left (bibliometrics) and right (IR)

A visual inspection of the terms "Information Retrieval" "Bibliometrics" and "Scientometrics" in

---

[1] See for example the annual conference European Conference on Information Retrieval (ECIR), and the bi-annual International Conference on Scientometrics and Informetrics (ISSI).

English Wikipedia with *Eyeplorer*[2] reveals different contextual spheres with "Information" "Database" "World-Wide Web" dominant for IR and "Science" "Bibliometrics" and "History of Science and Technology" for scientometrics. Despite of this difference efforts have been made to link the different communities. In 2013 only different workshops took place which intended to bring information retrieval (IR) and bibliometrics / scientometrics communities closer together, among them the workshop "Computational Scientometrics" held at iConference 2013[3] and 22nd ACM International Conference on Information and Knowledge Management (CIKM 2013)[4]. The guest editors of this special issue together with Howard D. White, Philipp Schaer and Peter Mutschke are responsible for another one, the workshop "Combining Bibliometrics and Information Retrieval"[5] held at the 14th International Conference of Scientometrics and Informetrics, Vienna, July 14-19, 2013. This workshop attracted more than 80 participants. The high interest among the bibliometricians was also generated by contributions from three Derek de Solla-Price-medal winners and leading-edge bibliometricians Michel Zitt, Wolfgang Glänzel and Howard D. White. An open call for contributions afterwards led to this issue.

**What exactly is now the relationship between Information Retrieval and Scientometrics?**
And why have those both fields apparently moved away from each other?

One factor for such a drift is the ongoing growth of the science system itself, with a hyperbolic growth of scholarly publications (Börner 2010, The Rise of Science and Technology page 4) and an accompanying diversification leading to an increasing number of specialities. The latter have been stated as having an usual size of 120-160 researcher (Kochen and Blaivas 1981), which probably has to do with a maximum size of a network with which one person still can pursuit a regular and indebt exchange of information. Another factor might be the amorphous nature of the overarching discipline of information science to which both areas belong. In a recent paper "Theoretical development of information science - a brief history" (Hjørland 2014) Hjørland quotes Howard D. White's testimony of 1999 "I see the field of library and information science (L&IS) as highly centrifugal and greatly in need of high-quality syntheses." (White 1999) Complementing such accounts, which mostly refer to West-European or North American authors, the inspection of Manfred Bonitz' (one of the information science pioneers in the East) personal library reveals analogous struggles in East to define the scope of information science and its relative position in the canon of scientific disciplines. Anecdotally, he reports about disputes among Soviet Union scientists after the publication of "The Foundation of Information Science"[6] by Michajlov, Chernyj and Giljarevskij (1968). One proposal was to subsume information science (as the science of scientific and technical information) under computer science (Bonitz 2010). This ambiguity in the definition of Information Science or Library and Information Science (LIS) penetrates also the relationship between Information Retrieval and Scientometrics.

---
[2] http://en.vionto.com/show/me/eyePlorer.com
[3] http://www.cse.unt.edu/~ccaragea/iConfWs-13.html
[4] http://www.cse.unt.edu/~ccaragea/CIKM-WS-13.html
[5] http://www.gesis.org/en/events/events-archive/conferences/issiworkshop2013/
[6] The Russian term in *Informatika*, in German Informatik - which is best been translated with Information Science. The English *Informatics* has a different meaning.

As the reader will see throughout the contributions to this special issue, the words *bibliometrics*, and *scientometrics*, sometimes even *informetrics* are used alternatively. They are related but by no means the same. According to the OECD Glossary of Statistical Terms "Bibliometrics is a statistical analysis of books, articles, or other publications."[7] while a reference is made to the Oxford Dictionary online 2013 (OECD Frascati Manual 2002). Indeed we find in the history of bibliometrics librarians as Bradford among the founders of the field. Wikipedia states that "Scientometrics is the study of measuring and analysing science, technology and innovation." But, at the same time on the article page for this term we read the editorial suggestion to merge the entry on scientometrics with the one on bibliometrics.[8] Scientometrics shares with bibliometrics the analysis of publications - but then in the area of science and technology. The parallel use of the different terms *bibliometrics, scientometrics,* and *informetrics* for the main conference series in this field hints to an on-going struggle of identity or more positively expressed to the on-going diversity of the field (Glänzel 2003; Cronin and Sugimoto 2014). While often used interchangeable, scientometrics usually is broader and also includes studies of expenditures, education, institutions, in short all metrics and indicators occurring in quantitative studies of the science system (De Bellis 2009). Informetrics is a bit different again, defined as study of the quantitative aspects of information (not only scientific information) (according to Wikipedia) or "all fundamental quantitative aspects of information science" as the journal of the same name states[9]. See also a bibliometric argument why informetrics is an own research field and needs a specific journal (Mayr and Umstätter 2007). Irene Wormell wrote about "informetrics" in the International encyclopedia of information and library science: "The field is becoming a scientific discipline that includes: all the statistical and mathematical analysis related to the study of information flows; evaluation of science and technology; library collection development; and documentation and information problems with strong links to the theoretical and methodological aspects of information retrieval." (Feather and Sturges 2003, p. 319).
Information retrieval is defined by Manning et al. (2008) as "finding material (usually documents) of an unstructured nature (usually text) that satisfies an information need from within large collections (usually stored on computers)". The paper of Wolfram in this issue reveals the complex and comprehensive nature of Information retrieval bridges from user studies over data models to the architecture of Information Systems. The field information retrieval with its main conferences SIGIR, TREC and ECIR is largely fragmented with a current trend to include more research from the Human Computer Interaction and Interactive IR sector.
The discussion during the ISSI workshop in Vienna highlighted the following features of

---

[7] http://stats.oecd.org/glossary/detail.asp?ID=198
[8] http://en.wikipedia.org/wiki/Scientometrics Accessed September 25, 2014 18:21
[9] http://www.journals.elsevier.com/journal-of-informetrics/

distinction:
- The **audiences** served by IR and scientometrics are very different. The former focuses on users of information systems as implemented on the web, in libraries and archives, the latter - in its form of evaluative bibliometrics - addresses primarily research manager and science policy makers in universities, funding agencies and ministries.
- With these different audiences come also different **goals** towards which algorithms of information processing used in both fields are tailored. IR supports an individual user to find paths through knowledge spaces. While devoted to an as good as possible match between search terms and materials in the collection at hand, serendipity and large coverage of the retrieved set of documents are not unwelcome features. For the evaluation of research groups the goal is to delineate the field of relevant works as sharp and precisely as possible. Again, we talk about a set of document retrieved. So far IR is at the beginning of any scientometric analysis. But its methodological core is the further analysis of the retrieved set of documents and the application of metrics and indicators. Any unintended extension of the retrieved set of core documents against which performance and impact of groups are mapped can lead to distortions in evaluative practices with all kind of consequences for institutional and individual careers. This explains why the struggle for 'good indicators' is fought with such a vibrance.
- The **scale and nature** of the collections or information spaces upon which IR or scientometrics/bibliometrics operate can be different. Information retrieval is not only applied for scholarly communication and related bibliographic databases. Its application area encompasses intelligence, business information, library catalogues, collections of musea and libraries, and the world-wide web as a whole (search engines). In contrast, scientometrics mainly operates in the world of journal articles, and only more recently opened towards scholarly communication on the web (webometrics and altmetrics).
- The **educational** paths towards IR and bibliometrics are different to some extent. This accounts further for the gap between IR and bibliometrics in fundamental research. In universities IR can be found as part both of Computer Sciences as well as of iSchools or Information Schools. Bibliometrics and Scientometrics are far less established in standard university curricula. However, as summer schools and professional training courses show, there exist a growing interest in this topic.[10]

As said before information retrieval is at the root of any scientometric or bibliometric study, and the research experience of authors in this special issue are spread over both areas. It is not a mere coincidence, as Karlsson et al. remember us, that the first event in the ISSI conference series hold the name *1st International Conference on Bibliometrics and Theoretical Aspects of Information Retrieval* (organized 1987 by Leo Egghe).[11] Returning to the aforementioned ambiguity in defining the field of information science Richard Smiraglia in a recent book abandons almost completely the notion of information science, uses information instead and analyses the history of the field using the lens of institutions and even more important through the practices of those working in this field, them being the real carriers of knowledge of the field (Smiraglia 2014).

---

[10] See e.g. the European Summer School for Scientometrics (since 2010) http://www.scientometrics-school.eu/ or the CWTS Course for Professionals "Measuring Science and Research Performance" http://www.cwts.nl/CWTS-Course-for-Professionals

[11] For more information about the ISSI please consult http://www.issi-society.org/past.html

This special issue invites the reader to do no less, to read the accounts of different scholars in the field, their perception of links between IR and scientometrics, and their identification of promising future lines of research and collaboration. For us as guest editors remains only to put out some own observation in this editorial to tease the reader to dig deeper into the contributions themselves.

That IR is much more than 'just' retrieval of a set of documents we learn from **Wolfram**'s paper (2015) "Information retrieval encompasses the processes of how information is represented, stored, accessed, and presented. IR systems represent implementations of these processes." **Glänzel** (2015) starts from the observation that bibliometrics needs a specific retrieval of documents, and further elaborates how bibliometric methods in turn can become an aid for information retrieval purposes. In a logical way he describe search strategies as sequences of decisions of inclusion and exclusion. He differentiates between unconditional and conditional search operations, leading to an elegant mathematical description of search strategy. The goal for such a search is to identify core documents or relevant documents. Using the schema proposed by Wolfram (Fig. 1) these questions seem to have most resonance with the *Retrieval processor*. **Zitt** (2015) in his turn discusses the consequences of different search strategies in the quest how to delineate a scientific field. He elaborates on two different ways of thinking - the a priori and the a posteriori perspectives in the process of performing retrieval or bibliometrics. The combination of lexical and reference based methods is discussed by Zitt as well as by Glänzel. The latter requires of course that references are part of *Information Database*. **Bar-Ilan** and **Levene**'s (2015) short paper points us to another kind of *Database*, the world wide web. They discuss how metrics developed in the realm of bibliometrics, the h-index, could be applied to other - much larger - information spaces.

Search strategies are tailored towards information needs, and those are different for different user groups. The paper of **Karlsson** and co-authors (2015) takes a step back and interrogates the meaning of relevance. What is relevant for whom, under which conditions? Relevance is also discussed by **White** (2015). He refers to the pragmatic side in information theory and links the relevance theory of Sperber and Wilson to a specific way to filter and eventually visually present information to a user. Karlsson and co-authors ask how the unavoidable amount of uncertainty can be made transparent but also practically managed. Information fusion - a term borrowed from the wider range of information science and knowledge management and staying for the combination of different information sources, could be one answer. This way - returning to Wolfram's more general scheme - the paper of Karlsson and co-authors combines aspects of *Users* with aspects of the information database available.

Maybe, it would be possible, to use Glänzel's mathematical description of a search strategy to express and maybe quantify the uncertainties resulting from the different nested selection criteria? Maybe it would be also possible to judge different *Retrieval Processors* as used in bibliometrics according to their uncertainty value? How could an expectation be quantified? How the different approaches could be benchmarked against such an expectation? What would be the ideal retrieval - the ideal delineation of a field? But maybe the last question in itself is slightly misleading given the variety of uses for IR techniques and audiences even for bibliometric analysis.

**Mutschke** and **Mayr** (2015) (their paper in this issue is an extension of the ideas developed and evaluated in Mutschke et al. 2011), and **Abbasi** and **Frommholz** (2015) explore the use of regularities in scientific communication, as depicted in models of science dynamics, for IR. If we know that authors or papers central in networks of co-authorship, co-word, or co-references seem to have a lighthouse function for scientific communication (thus the Matthew effect in Science according to Merton) would it than not be appropriate to use those insights to provide user with recommendations? If we also know from bibliometrics (Zitt, Glänzel) that different metrics can depict the core of a field, but also weak links to other fields, could this not be presented to an user in order to allow her to triangulate between different perspectives. Abbasi and Frommholz propose one such solution to tackle uncertainty by offering browsing strategies in interactive user interfaces based on poly-representation of sets of relevant results.

Experiences with *Interfaces* is one advantage of IR above bibliometrics. The importance of interfaces is also visible in White's contribution. Constructing and testing different interfaces is one way to give authority back to the user. Not so much a-priori reduction of uncertainty is the goal here, but means to make ambiguity and uncertainty productive. This would also require to order and mark the different ways to navigate through a selection, and logical models as the one proposed by Glänzel could be made transparent in an interactive search.

An alliance between bibliometrics and IR concerning public information databases could also support a much needed benchmarking for bibliometric algorithms. The latter is strangulated by licence issue concerning databases of commercial information providers. This could also help to organize a better knowledge exchange between IR and bibliometrics. The experience of bibliometrics in exploring all kind of different possible correlations between the (co-)occurrence of different elements in the bibliographic record (and beyond - see altmetrics) has by no means sufficiently found its way into explorative IR interfaces in collections. Here IR can learn a lot from bibliometrics. In turn, IR can offer an in-vivo, driven by user counter-test of what is thought to be the way along which scientific information diffuses. Based on such envisioned user studies one could think of new models of science, which not only include structures visible in the formal scholarly communication but also the ephemeral but equally important, bending ways in which humans acquire information and build new knowledge.